\documentclass[11pt,a4paper]{article}

\usepackage[T1]{fontenc}
\usepackage[utf8]{inputenc}
\usepackage[margin=2.0cm]{geometry}
\usepackage{amsmath}
\usepackage{amssymb}
\usepackage{booktabs}
\usepackage{graphicx}
\usepackage{longtable}
\usepackage[font={small,sf},labelfont=bf,skip=4pt]{caption}
\usepackage{flafter}
\usepackage{float}
\usepackage{xcolor}
\usepackage[numbers,sort&compress]{natbib}
\usepackage[colorlinks=true,linkcolor=blue!50!black,citecolor=blue!50!black,urlcolor=blue!50!black]{hyperref}

\graphicspath{{figures/}}
\newcommand{\benchname}{Report2Alpha-Bench}

\title{\benchname: Evaluating LLM Agents on\\End-to-End Quantitative Research}
\author{
  Yun Zhang$^{1,*}$\qquad Hui Yang$^{1}$\qquad Wei Meng$^{1}$\\[6pt]
  {\normalsize $^{1}$Fundora Lab, \url{https://fundora.red/}}\\[2pt]
  {\normalsize $^{*}$Corresponding author: \href{mailto:yzhyun0321112x@gmail.com}{yzhyun0321112x@gmail.com}}
}
\date{November 2026}

\begin{document}

\maketitle

\begin{abstract}
We evaluate whether LLM agents can independently carry out a several-hour quantitative research project: read a brokerage research report, make a plan, and over 7 rounds of iteration propose hypotheses, write factor code and improve it from backtest feedback, seeing only 2022--2023 market data.
Like Terminal-Bench-Science, the task evaluates overall research ability; unlike it, there is no directly checkable unique answer: the model must repeatedly propose hypotheses on very noisy data, revise its understanding from experiments, disentangle multiple factors and design ablations without overfitting in-sample, much like complex real-world decisions with many interacting factors.
After the research, a fixed rule using only 2022--2023 results selects a low-correlated \emph{factor pool} from all candidates the model produced, and scores it on unseen 2024--2025 data after removing common style returns such as size and momentum; the same rule on round-0 candidates alone gives a report baseline that ``implements the report without iterating''.
We run gpt-6.1-sol (reasoning effort xhigh), gpt-6.1-terra (xhigh), muse-spark-1.3 (xhigh) and deepseek-v4.1-flash (max) once on each of 7 reports. Of the 28 runs, 19 factor pools pass the usability threshold; the average score is 48.4 (0--100 scale), against 25.4 for the report baseline and 35.9 for the models' own picks.
From similar report baselines (17.0--31.0), the models separate after iteration to 74.7 (gpt-6.1-sol), 56.6 (gpt-6.1-terra), 35.8 (deepseek-v4.1-flash) and 26.2 (muse-spark-1.3); their candidates look equally good in training and diverge only in testing, with the leaders proposing more diverse ideas per round and ending with more effective factors in the pool.
Evaluating research agents thus requires scoring on frozen future data after removing styles, and measuring ``what the research produced'' separately from ``what was finally picked''.

\end{abstract}

\section{Introduction}
\label{sec:intro}

\begin{figure}[t]
  \centering
  \includegraphics[width=\linewidth]{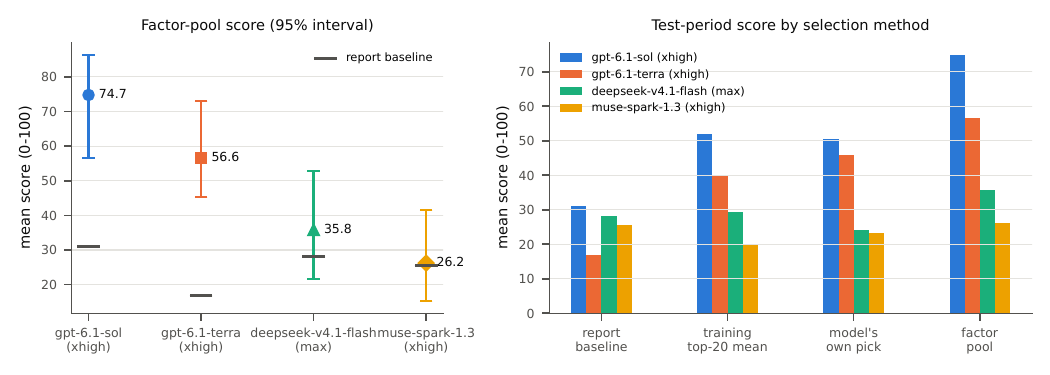}
  \caption{Main results. Left: average factor-pool scores (0--100 scale) of the 4 models, with reasoning effort in parentheses; error bars are 95\% intervals obtained by resampling trading days, and short dashes mark the average of the report baseline. Right: average test-period scores of four selection methods; see Table~\ref{tab:baselines} for definitions.}
  \label{fig:headline}
\end{figure}

LLM agents can already work in a terminal for hours at a time, completing tasks such as software engineering, machine learning experiments and paper reproduction \citep{merrill2026terminalbench,jimenez2024swebench,chan2025mlebench,starace2025paperbench}.
Once the task changes from ``answering a question'' to ``completing a piece of research'', evaluation becomes harder: research output usually has no single correct answer, and scoring it with human or LLM reviewers is costly and hard to reproduce.
We need tasks whose process is open-ended and requires long-horizon planning, but whose results can still be scored objectively and automatically. Quantitative stock-selection research fits: its ``ground truth'' is future market data itself.

In our task, the model starts from a brokerage research report and turns its ideas into a program that scores every stock every day (a \emph{factor}); over 7 rounds of iteration it proposes new hypotheses from backtest feedback and improves.
The report is only a starting point: we do not ask for a faithful reproduction, but whether the model can build on the report's core logic and iterate its way to factors that still work on future data.

The task evaluates more than a one-time translation from prose to code; it requires a sustained research loop: propose a hypothesis, run an experiment, deepen the understanding of the alpha logic from the result, and propose the next hypotheses. The research must iterate in sample without overfitting to in-sample results; it cannot rely on a single summary metric, but must sort out the joint effects of several factors, tell which effects come from the logic and which from noise, and use ablations to determine which conditions really matter. It therefore requires long-horizon planning, implementing code to specification, proposing hypotheses and extending them from results, disentangling the effects of multiple factors, and designing ablations. Many real-world decisions face similarly noisy data with many influencing factors; we conjecture that models that do well on this task may also be more reliable on such decisions, but this paper does not test that.

Scoring this kind of research on historical market data has three difficulties. The first is overfitting: repeated trial and error on the same data will always turn up factors that work by chance \citep{bailey2014deflated,harvey2016cross}, so we split the data into a training period and a test period, and the model sees only the former during research. The second is style bets: a factor that leans toward small caps or recent winners may earn a high return in one year by chance and give it all back when the market changes, so we remove the returns explained by common styles before scoring and set a threshold so that only results usable in practice earn points. The third is the luck of picking: one research run produces hundreds of candidate factors, and if the model picks its own submission, the score also reflects how well it picked, so we instead use a fixed rule that looks only at training-period results to select the factors to be scored (the \emph{factor pool}), measuring ``what the research produced'' separately from ``what was finally picked''.

We evaluate four models, gpt-6.1-sol, gpt-6.1-terra, deepseek-v4.1-flash and muse-spark-1.3; each completes research once on each of 7 reports, 28 runs in total, with scores on a 0--100 scale (Figure~\ref{fig:headline}). We have three main findings.
\begin{enumerate}
  \item The models can produce usable factors. 19 of the 28 factor pools pass on the test period, averaging 48.4, above the report baseline that uses only round-0 candidates (25.4) and above the submissions the models picked themselves (35.9) (\S\ref{sec:overall}).
  \item The gap forms during iteration. The four models' report baselines are similar; after iteration they separate to 74.7, 56.6, 35.8 and 26.2. Each model's candidates perform similarly on the training period and diverge only on the test period (\S\ref{sec:gap}, \S\ref{sec:stage}).
  \item The gap is associated with the diversity of ideas. The leading models propose more diverse ideas per round, keep more low-correlated effective factors in the pool, and fall back less on the test period; neither reading the report nor writing code is a major source of the gap, and engineering failures are concentrated in a few runs (\S\ref{sec:analysis}). In addition, spending more tokens does not mean doing better research: gpt-6.1-terra uses about half the tokens of gpt-6.1-sol and scores second only to it, while deepseek-v4.1-flash uses about 2.7 times as many and scores less than half as much (\S\ref{sec:efficiency}).
\end{enumerate}

Our contributions are:
\begin{itemize}
  \item a task and environment for evaluating end-to-end quantitative research by LLMs: real reports as input, executable factor code as output, deterministic backtests on market data the model has never seen (\S\ref{sec:task});
  \item a scoring method that measures research output separately from final picking, including the factor pool, the report baseline and control groups, ability-based process metrics and confidence intervals (\S\ref{sec:method});
  \item a quantitative analysis of where the gaps between models come from, with traceable cases (\S\ref{sec:results}, \S\ref{sec:analysis}).
\end{itemize}

\section{Related Work}
\label{sec:related}
Among financial benchmarks, FinQA \citep{chen2021finqa}, FinanceBench \citep{islam2023financebench} and FinBen \citep{xie2024finben} examine question answering and numerical reasoning; StockBench \citep{stockbench2025} examines trading decisions; ILE \citep{ile2026} evaluates fundamental research under a frozen information set, but relies on rubrics and an LLM judge.
FinForesight \citep{liu2026finforesight} compares evaluation within the pre-training period, after the training cutoff and in real time, and finds that backtest scores within the pre-training period are generally inflated, which suggests that historical backtests may mistake memorization for prediction.
None of these benchmarks asks the model to hand in research outputs that can be run and objectively tested on future data.

In factor mining, AlphaGen \citep{yu2023alphagen}, Alpha-GPT \citep{wang2023alphagpt}, R\&D-Agent-Quant \citep{li2025rdagentq}, AlphaAgent \citep{tang2025alphaagent}, CogAlpha \citep{liu2025cogalpha} and X-Alpha \citep{liu2026xalpha} generate factors automatically with reinforcement learning or LLMs; X-Alpha also distills research paths from research reports.
These works aim to mine good factors and compare different methods; we do the reverse, fixing the research workflow and comparing the research ability of different models.

Among general agent benchmarks, SWE-bench \citep{jimenez2024swebench}, MLE-bench \citep{chan2025mlebench}, PaperBench \citep{starace2025paperbench} and Terminal-Bench \citep{merrill2026terminalbench} evaluate tasks that require long stretches of continuous work. They check whether tests pass or whether a metric reaches a target, and usually do not treat historical overfitting in financial backtests as a primary evaluation risk.
Terminal-Bench-Science (TB-Science) \citep{tbscience2026} goes further and collects expert-curated research-workflow tasks from the life, physical, earth, mathematical and engineering sciences, and requires results to be objectively verifiable in the terminal.
Like TB-Science, our benchmark evaluates overall research ability, and its task format follows the Harbor framework \citep{harbor2025} that TB-Science uses; the difference lies in the research process itself. The TB-Science project description emphasizes objective verification in the terminal, whereas quantitative research in this benchmark has no unique correct answer that can be checked directly: the model must repeatedly propose hypotheses, read experimental results, deepen its understanding of the alpha logic and propose new hypotheses, while guarding against overfitting to in-sample results, not relying on a single metric, and telling which parts of a result come from the logic and which from noise. Such tasks share features with complex real-world decisions involving noisy data and many influencing factors, so for scoring we separate a training period from a test period and remove returns due to styles. Table~\ref{tab:related} summarizes how our benchmark differs from the closest existing ones.

\begin{table}[htbp]
  \centering
  \caption{Comparison with existing benchmarks. The ``agent loop'' and ``information cutoff control'' entries for StockBench and ILE are taken from the comparison table of ILE \citep{ile2026}; the TB-Science row follows its project description \citep{tbscience2026}; ``--'' means not reported.}
  \label{tab:related}
  \footnotesize
  \setlength{\tabcolsep}{3pt}
  \resizebox{\linewidth}{!}{\begin{tabular}{@{}llccllll@{}}
    \toprule
    Benchmark & Object of evaluation & Agent loop & Executable output & Scoring & Information cutoff control & Process measurement & Confidence interval \\
    \midrule
    StockBench & Stock trading decisions & Yes & -- & Trading return & Yes & -- & -- \\
    ILE & Full fundamental research workflow & Yes & -- & Rubrics and LLM judge & Yes & Yes & -- \\
    Terminal-Bench & Terminal tasks & Yes & Yes & Test scripts & N/A & Trajectory error analysis & Yes \\
    TB-Science & Scientific research workflows & Yes & -- & Objective verification in the terminal & -- & -- & -- \\
    \benchname & Full quantitative research workflow & Yes & Yes & Deterministic backtest on new market data & Partial (workflow convention) & Yes & Yes \\
    \bottomrule
  \end{tabular}}
\end{table}

\section{Evaluation Task}
\label{sec:task}

This section describes the task from the model's point of view: what the model is given, what it has to do, and which decisions are left to the model.

\subsection{Inputs and Outputs}
\label{sec:io}
The model is given three things: a brokerage financial engineering research report (dozens of pages, with formulas and charts), a read-only copy of historical market data, and a set of callable compilation and backtesting tools.
The data include daily bars, 5-minute and 1-minute bars; indicators computed from tick-by-tick trades, tick-by-tick orders and order-book snapshots (provided only after aggregation to 1-minute or 5-minute intervals, with no raw tick records); and the style and industry exposures of every stock from the Barra CNE5 model, 1,454 fields in total. For the 7 reports in this benchmark, these fields are enough to cover all kinds of underlying data their factors need, so the model is not prevented by missing data from reproducing a report or improving on it. Research and scoring both use only the tradable stocks of the Shanghai and Shenzhen markets, excluding ST stocks and stocks listed for less than one year.

A \emph{factor} is a score assigned to every stock every day to predict its future return relative to other stocks (see Appendix~\ref{app:glossary} for terminology). Factors are written in C++ and computed in a framework that replays historical data day by day and every 5 minutes.
A factor is computed at the open of each trading day and may only use data already known at that time; using data that only become available after that day's open (such as the day's close) is called \emph{look-ahead} and counts as cheating.

The model's output is every candidate factor it implemented and backtested during research. At scoring time the model does not do the picking; a fixed rule selects a factor pool from these candidates (\S\ref{sec:pool}), and the submission the model picks itself at the end serves only as a control.

\subsection{Research Workflow}
\label{sec:work}
\begin{figure}[htbp]
  \centering
  \includegraphics[width=\linewidth]{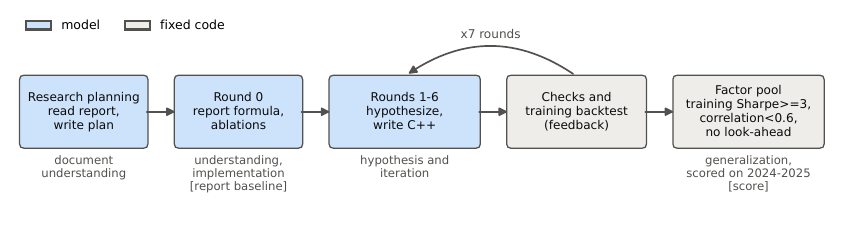}
  \caption{Research workflow. Blue marks model steps and gray marks fixed code; below each stage is the ability or scoring object it corresponds to.}
  \label{fig:workflow}
\end{figure}

The model's work can be summarized in three steps (Figure~\ref{fig:workflow}):
\begin{enumerate}
  \item Understand the report and make a plan: identify the report's core idea and original method, and decide on research directions;
  \item Iterate: over 7 rounds, repeatedly ``propose a hypothesis $\rightarrow$ implement $\rightarrow$ read feedback $\rightarrow$ adjust''. Round 0 usually includes the report formula, controls that change only one thing (\emph{ablations}: remove or replace one component and keep the rest) and a few new directions;
  \item Pick a submission: at the end of the research, the model picks its final submission from its own candidates following the criteria given in the prompt. This step serves only as a control; the official score comes from the factor pool selected by the fixed rule.
\end{enumerate}
Each \emph{hypothesis} corresponds to one piece of factor code; once it passes the engineering checks and the training-period backtest it becomes a \emph{candidate}.
Engineering checks, backtests and feedback are all done by fixed code, not by the model. The \emph{engineering checks} cover compilation, a trial run, coverage (the share of stocks with a factor value each day) and cross-sectional degeneracy (whether all stocks have exactly the same factor value on a day); if a check fails, the model must fix the code and resubmit within the same session, with at most 3 fixes.
Each round allows at most 100 hypotheses; the next round receives only the feedback of the latest round and an index of historical files, and the model has to look up earlier details itself.
During research the model can only see backtest results for 2022--2023 (the \emph{training period}); scoring uses 2024--2025 (the \emph{test period}), which the model cannot see.

One run lasts 1.6--7.9 hours and opens 24 model sessions (research plan, plan packaging, reflection and coding in each of the 7 rounds, and the final pick), producing hundreds to over a thousand tool calls.

\subsection{Stage Prompts}
\label{sec:prompts}
Each model step has a system prompt that states what the stage does, which materials it may read and what format it must deliver (Table~\ref{tab:prompts}). The prompts encode researchers' experience in factor research; their main purpose is to keep the iteration disciplined and to reduce overfitting to the training period.

The longest prompt, and the one that constrains research method the most, belongs to the reflection stage (about 26,000 Chinese characters). Its core is a cumulative iteration tree maintained by the model itself: each candidate records its ID, its parent factors and the research change made in this round; each round first copies the previous round's tree and then appends new nodes, keeping failed and refuted nodes as well, so that the next round decides on the full history rather than only on the latest feedback.
The prompt asks the model to first confirm that an experiment is valid, then read the full ten-group returns, propose several explanations that can be told apart, and write the next step as a falsifiable prediction; whether a component is necessary is tested with controls that change only that component and keep the rest fixed (ablations). Progress in logic and progress in performance are assessed separately, and a rise in performance alone does not count as confirming a hypothesis. The prompt also repeatedly reminds the model that all feedback comes from the training period: small differences between sibling factors that differ only in a window or threshold are not a reason to switch parents, and the model should prefer parameter regions where neighbouring settings perform alike rather than chase a single best point.
The prompt of the final-pick stage takes the same stance: whether a version qualifies is judged mainly on logic (80\% weight) and secondarily on performance (20\%), stable performance matters more than peak performance, versions with fewer parameters are preferred, and correlation between versions decides whether to keep one or several.

While building the framework, researchers observed in their tests that without these constraints models (for example deepseek-v4.1-flash) were more prone to iterate after the best training-period score and overfit in sample. This observation was not tested with a controlled experiment in our 28 runs; the effect of the prompts on scores is left for future work (\S\ref{sec:limitations}).

\begin{table}[htbp]
  \centering
  \caption{Main requirements of the prompt for each model step (plan packaging only reformats the plan and is not listed).}
  \label{tab:prompts}
  \small
  \begin{tabular}{@{}p{2.2cm}p{12.3cm}@{}}
    \toprule
    Stage & Main requirements \\
    \midrule
    Research plan & Read the full report text and figures; extract the original factors, formulas and parameters with their locations; propose implementations of the report formulas where possible; keep research directions within the phenomena and mechanisms of the report and their direct extensions; write no code, and do not present anything absent from the report as its definition. \\
    Reflection & Maintain the cumulative iteration tree; confirm experiment validity, then read the full group returns and propose distinguishable explanations and falsifiable predictions; test whether components are necessary with controls (ablations); assess logic and performance separately; feedback is training-period only, so do not chase a single best point; at most 100 hypotheses per round, with no quota. \\
    Coding & Follow the logic written in the reflection stage, map its variables to real data fields (an approximate field with the difference noted when no exact match exists), and implement it with the factor-development guide (skill); for factors with parents, copy the parent code before modifying it; deliver only after a successful build. \\
    Final pick & Read only the last round's iteration tree and training-period performance; logic first (80\%), performance second (20\%); stability before peak, fewer parameters preferred; correlation decides how many versions to keep. A control only. \\
    \bottomrule
  \end{tabular}
\end{table}

The factor-development guide used in the coding stage covers the data interface, the available computation functions and the rules on ``which data may be used at which moment''. With this guide, even the lightweight deepseek-v4.1-flash reliably writes code that runs: for 3 of the 4 models at most 0.6\% of hypotheses fail the engineering checks (\S\ref{sec:reliability}). The benchmark therefore separates models mainly by their ability to reflect and propose hypotheses, not by their ability to write code.

Table~\ref{tab:scaffold} distinguishes three kinds of components in the workflow; this paper evaluates the ``model-autonomous'' part.

\begin{table}[htbp]
  \centering
  \caption{Three kinds of components in the workflow.}
  \label{tab:scaffold}
  \small
  \begin{tabular}{@{}p{2.4cm}p{12.1cm}@{}}
    \toprule
    Code-enforced & Stage order, the 7-round loop, engineering checks, training-period backtests and feedback, the factor-pool rule and scoring. \\
    Prompt-guided & The structure of the plan; the research method of the reflection stage; the coding guide (skill); the criteria for the model's final pick (a control only) (\S\ref{sec:prompts}). \\
    Model-autonomous & How to understand the report, which hypotheses to propose, how many per round, which existing factor to modify, how to map the variables in the report to data fields, how to implement. \\
    \bottomrule
  \end{tabular}
\end{table}

\subsection{Task Set}
\label{sec:taskset}
The task set contains 7 reports, all published before the test period (Table~\ref{tab:tasks}), chosen from 431 price--volume and order-flow reports that run through the whole engineering pipeline; the task set can be extended with the same method.
At selection time, we looked at how earlier automated research runs on these reports in this repository performed on the training period, divided the reports into four difficulty tiers, and chose 1--2 reports per tier to cover different types of reports: \emph{improvable by research} (A, the report formula is weak but improves clearly after iteration), \emph{strong baseline} (B, implementing the report formula as written is already good), \emph{high ceiling} (C, further gains are possible on top of a strong baseline) and \emph{hard} (D, still weak after iteration); see Appendix~\ref{app:taskset} for details.

\begin{table}[htbp]
  \centering
  \caption{Task set. The first letter of the ID is the difficulty tier at selection time: A improvable by research, B strong baseline, C high ceiling, D hard. Formula completeness is an annotation result: ``complete'' means the core factor is given with directly implementable formulas and parameters, and ``missing parameters'' means the structure is given but key parameters or definitions are missing.}
  \label{tab:tasks}
  \small
  \setlength{\tabcolsep}{4pt}
  \begin{tabular}{@{}lp{4.4cm}p{2.1cm}lp{2.2cm}p{2.0cm}@{}}
    \toprule
    ID & Topic & Broker & Published & Main data & Formula completeness \\
    \midrule
    A1 & Identifying stock-level momentum effects and combining them into a ``team coin'' factor & Founder Securities & 2022-06-11 & daily, minute, tick & missing parameters \\
    A2 & Identifying information shocks from stock price jumps & Zheshang Securities & 2021-02-20 & daily, minute & complete \\
    B1 & Price behavior at moments of volume surges & Founder Securities & 2022-04-12 & daily, minute & complete \\
    B2 & The interaction between stock prices and money flows & Huaan Securities & 2023-06-11 & daily, minute, tick & missing parameters \\
    C1 & Characterizing large-money behavior by minute-level average trade amount & Kaiyuan Securities & 2022-06-12 & daily, minute & complete \\
    C2 & A misaligned version of price--volume autocorrelation & Soochow Securities & 2021-03-01 & daily, minute, order book & complete \\
    D1 & Searching for similar candlestick patterns & Dongguan Securities & 2020-07-28 & daily & missing parameters \\
    \bottomrule
  \end{tabular}
\end{table}

\section{Evaluation Method}
\label{sec:method}

This section explains how one research run is scored: first the scoring formula, then the rule that selects the factors to be scored and the controls we construct, and finally process measurement, anti-cheating measures and confidence intervals.

\subsection{Outcome Scoring}
\label{sec:outcome}
Scoring replays market data from 2024--2025 (485 trading days). This period is later than the publication dates of all reports, and the model cannot see backtest results for it during research.

Scoring should answer ``how strong a stock-selection signal the model found'', not ``which market style did it happen to bet on''. In 2024 large and small caps took turns leading the A-share market, and the 2025 rally rewarded momentum stocks that chased recent winners; a factor that merely leans toward one style can earn a large return in a single year.
Scoring therefore first applies a cross-sectional nonlinear style neutralization to the factor values, thoroughly removing their association with styles such as size and momentum; returns are then measured as Barra-based \emph{style-free returns}, i.e.\ what remains after regressing the returns of all stocks on the market, 32 industries and 10 Barra styles once per day. Results from different years thus mainly reflect stock-selection ability itself.

After removing styles, each day stocks are split into ten equal groups by factor value, and we compute the annualized Sharpe ratio $S$ (mean return divided by volatility, annualized; higher is better) of the style-free return of the five highest groups over the average of all ten \citep{sharpe1966,sharpe1994}.
$S<2$ scores 0, otherwise the score is $\min(S/6,1)$, multiplied by 100; $S\ge2$ is called a \emph{pass}.
The threshold of 2 comes from practitioners' experience: factors with $S$ below 2 have volatile performance and are hard to use as alpha factors in live trading. It is also hard to reach by chance: over the 485 test days, $S=2$ corresponds roughly to a $t$-statistic of 2.79 \citep{lo2002sharpe}.
Scoring is a deterministic backtest with no LLM judge. Multiple factors are first combined into one signal, weighted by each factor's performance over the past 60 days, and then scored; formulas are in Appendix~\ref{app:scoring}.

\subsection{Factor Pool and Controls}
\label{sec:pool}
The factors to be scored are not picked by the model but selected by a fixed rule based only on training-period results; we call them the \emph{factor pool}:
\begin{enumerate}
  \item all candidates are sorted by training-period $S$ from high to low, and those with $S<3$ are excluded;
  \item candidates are examined in turn, and one enters the pool only if its correlation of training-period daily returns with every factor already in the pool is below 0.6;
  \item factors with look-ahead are removed and their places are not refilled; the pool is replayed on the test period to give the score of the run, and an empty pool scores 0.
\end{enumerate}
The entry threshold of 3 is above the scoring threshold of 2, which partly offsets the optimism from repeated trial and error; the correlation constraint makes the pool measure ``how many distinct effective signals were found'' rather than many variants of one signal. Details of the rule are in Appendix~\ref{app:scoring}.

To see at which step the gap appears, we build several controls from the same candidates (Table~\ref{tab:baselines}) without rerunning any model. The most important is the \emph{report baseline}: the same rule applied to round-0 candidates only, representing what can be achieved ``starting from the report without iterating''; its difference from the factor pool is the gain from 6 rounds of iteration.
The submission the model picks itself is also kept as a control: its difference from the pool reflects the difference between two scoring protocols, not a measure of model ability.
Every control is assembled into a submission in the same format as the pool and replayed by the same scoring container, 575 distinct factor sets in total.
``Training-period $S$'' here is computed after the fact with the scoring formula on the training period; in its feedback the model sees the return and Sharpe ratio of each of the ten groups, turnover and quarterly performance, but not this number itself.

\begin{table}[htbp]
  \centering
  \caption{Scored object and controls. Each run gets one value for each.}
  \label{tab:baselines}
  \small
  \begin{tabular}{@{}lp{7.2cm}p{4.6cm}@{}}
    \toprule
    Name & Definition & Question answered \\
    \midrule
    Factor pool & Pool selected from all candidates by the rule above (the score) & What the research finally produced \\
    Report baseline & The same rule applied to round-0 candidates only & How much iteration adds on top of the report \\
    Own pick & The submission the model picks itself at the end of the research, following the prompt's criteria & Fixed rule versus the model's own picking \\
    Best in training & The single candidate with the highest training-period $S$ among all candidates & What betting on a single factor achieves \\
    Training top-20 mean & The mean of the 20 candidates with the highest training-period $S$, each scored on its own (not an achievable strategy) & Test-period quality of the candidates as a whole \\
    \bottomrule
  \end{tabular}
\end{table}

\subsection{Process Measurement}
\label{sec:process}
The final score alone cannot tell us ``where the difference lies''. We divide the abilities a research run needs into five---document understanding, implementation reliability, hypothesis and iteration, generalization, and long-horizon stability---and define a set of process metrics for each from the run records (Table~\ref{tab:process}). \emph{Generalization} refers to how much of the candidates that look good on the training period is left on the test period; time and tokens are reported separately as efficiency. Computation methods and annotation rules are in Appendix~\ref{app:annotation}.

\begin{table}[htbp]
  \centering
  \caption{Process metrics defined by ability. ``Annotation'' means a judgment made by reading the run artifacts according to written rules (Appendix~\ref{app:annotation}).}
  \label{tab:process}
  \small
  \begin{tabular}{@{}p{2.6cm}p{8.7cm}p{3.0cm}@{}}
    \toprule
    Ability & Metric & Source \\
    \midrule
    Document understanding & Plan recall of the report's core methods & Annotation \\
    Implementation reliability & Share of hypotheses that fail engineering checks; counts of engineering check failures and of resubmissions with unchanged source & Check logs \\
    Hypothesis and iteration & Hypotheses per round; average correlation of training-period daily returns among candidates in the same round (\emph{same-round correlation}; lower means more different ideas); share of improved factors whose training-period $S$ is higher than that of all their parents; number of factors in the pool & Run records, training backtests \\
    Generalization & Train--test gap (mean training-period $S$ of the pool's factors minus the pool's test-period $S$); count of training-period overfitting (gap $\ge1.5$ and pool test-period $S<2$) & Training backtests, scorer \\
    Long-horizon stability & Counts of rewrites requested because of unqualified output; whether the research goal or records drift after context compaction & Session records, annotation \\
    \bottomrule
  \end{tabular}
\end{table}

\subsection{Anti-Cheating and Confidence Intervals}
\label{sec:integrity}
The split into training and test periods and the pool entry threshold address overfitting; the model may also draw on market data memorized during pre-training \citep{glasserman2023lookahead}, which we discuss in Section~\ref{sec:limitations}.
This subsection is concerned with whether the research process itself crossed any line. Table~\ref{tab:isolation} lists the isolation measures between research and scoring: ``mechanism'' means that any violation makes the run fail; ``convention'' means the measure is enforced by prompts and the workflow and confirmed by after-the-fact checks.

\begin{table}[htbp]
  \centering
  \caption{Isolation measures and how they are checked. See Appendix~\ref{app:annotation} for the look-ahead check.}
  \label{tab:isolation}
  \small
  \begin{tabular}{@{}p{5.6cm}lp{6.6cm}@{}}
    \toprule
    Measure & Type & How it is checked \\
    \midrule
    Root file system, market data and reports are read-only; the pool program and the scorer are offline, only read research artifacts and recompile them; reference answers do not enter the research environment & Mechanism & Mounts checked at startup; scoring container tested to be read-only, offline and without system privileges \\
    Test-period market data must not be read & Convention & Run-log search: no command reading test-period market data found \\
    No network access & Convention & Run-log search: no web search, download or installation \\
    The scoring configuration must not be used for tuning & Convention & Run-log search: 1 read of the scoring configuration, whose content matches information public in the prompts \\
    Factors must not use look-ahead & Mechanism & Static source check with call-chain tracing; factors with look-ahead do not enter the pool; of the 3,897 candidates only the 210 of C1-muse read the day's close, so its pool is empty \\
    \bottomrule
  \end{tabular}
\end{table}

Since scoring depends only on market replay, we can resample the trading days of the test period in 20-day blocks to give 95\% intervals for average scores and score differences. Gaps between models are tested in two ways: by comparing average scores on the same resampled trading days, and by comparing the 7 tasks one by one (Appendix~\ref{app:scoring}).

\section{Experimental Setup}
\label{sec:setup}
Table~\ref{tab:models} gives the configurations of the 4 models. Every model call in a run uses the same model and reasoning effort, so that the score can be attributed to a single model. Each model runs once on each task, 28 runs in total. The models work in a terminal through Codex CLI \citep{openai2025codexcli}: the two GPT models support this toolchain natively and the other two are connected through a compatible configuration, so the conclusions are limited to performance under this toolchain. All test-period results are replayed after the research and were not used during it. In cases and tables below we abbreviate the models as sol, terra, deepseek and muse.

\begin{table}[htbp]
  \centering
  \caption{Configurations of the evaluated models. Context windows are taken from the model catalog or the model list cached at run time.}
  \label{tab:models}
  \small
  \begin{tabular}{@{}llll@{}}
    \toprule
    Access name & Reasoning effort & Access method & Context window \\
    \midrule
    gpt-6.1-sol & xhigh & Codex CLI native & 500K \\
    gpt-6.1-terra & xhigh & Codex CLI native & 500K \\
    deepseek-v4.1-flash & max & Codex CLI, compatible config & 1M \\
    muse-spark-1.3 & xhigh & Codex CLI, compatible config & 272K \\
    \bottomrule
  \end{tabular}
\end{table}

\section{Results}
\label{sec:results}
This section reports what we observe: whether the models can produce usable factors, how large the gaps between them are, at which step the gaps appear, and efficiency. Explanations are discussed in Section~\ref{sec:analysis}.

\subsection{Usable Factors}
\label{sec:overall}
Most factor pools pass on the test period, clearly above the report baseline. 19 of the 28 factor pools pass, averaging 48.4 points; the report baseline averages 25.4 points with 12 passes (Table~\ref{tab:baselines-main}).
The 95\% interval of how much the pool scores above the report baseline is [16.9, 31.6] (Table~\ref{tab:pairwise}). Each run has 7 rounds (rounds 0--6): round 0 starts from the report formulas and the next 6 rounds iterate. The pools contain 106 factors in total, 90 of which come from the iterations in rounds 1--6 and only 16 from round 0: the improvement comes mainly from iteration on top of the report, not from implementing the report itself.

The pool selected by the fixed rule is also above the submissions the models picked themselves. The models' own picks average 35.9 points, and the pool scores [5.7, 15.9] above them; taking only the single factor with the best training-period performance gives 41.0 points, also below the pool. Only 6 of the 44 factors the models picked themselves enter the pool.
This comparison concerns two scoring protocols, not a measure of model ability: the picking objective the prompt gave the models during research included a turnover penalty that differs from the scoring criterion, which works against the models' own picks. What it supports is that scoring by a fixed rule reflects what the research produced better than scoring the model's final pick.

\begin{table}[htbp]
  \centering
  \caption{Test-period scores of the factor pool and each control (0--100, mean over 7 tasks; passing tasks in parentheses). Empty pools score 0.}
  \label{tab:baselines-main}
  \small
  \setlength{\tabcolsep}{4pt}
  \begin{tabular}{@{}lccccc@{}}
    \toprule
    & gpt-6.1-sol & gpt-6.1-terra & deepseek-v4.1-flash & muse-spark-1.3 & Mean \\
    \midrule
    Report baseline & 31.0 (3) & 17.0 (2) & 28.1 (4) & 25.5 (3) & 25.4 \\
    Own pick & 50.5 (5) & 45.8 (5) & 24.0 (3) & 23.1 (2) & 35.9 \\
    Best in training & 64.3 (7) & 45.7 (5) & 32.5 (4) & 21.2 (2) & 41.0 \\
    Training top-20 mean & 51.9 & 40.1 & 29.5 & 19.7 & 35.3 \\
    \textbf{Factor pool} & \textbf{74.7 (7)} & \textbf{56.6 (5)} & \textbf{35.8 (4)} & \textbf{26.2 (3)} & \textbf{48.4} \\
    \midrule
    Mean factors in pool & 5.7 & 3.7 & 3.1 & 2.6 & 3.8 \\
    \bottomrule
  \end{tabular}
\end{table}

\subsection{Differences Between Models}
\label{sec:gap}
The four models are clearly separated. gpt-6.1-sol passes all 7 tasks with an average of 74.7; it is followed by gpt-6.1-terra (56.6), deepseek-v4.1-flash (35.8) and muse-spark-1.3 (26.2).
On resampled trading days, 5 of the 6 pairs of models differ significantly, the exception being deepseek versus muse; comparing task by task has limited power with only 7 tasks, and only sol versus muse reaches the 0.05 level (Table~\ref{tab:pairwise}). Removing C1 and D1, the two tasks with the largest gaps, leaves the ranking unchanged (69.4, 49.3, 39.4 and 36.7 points).

The difficulty of the task itself also matters a great deal. Task by task (Table~\ref{tab:per-task}), all 4 models pass B1 and C2, while only gpt-6.1-sol passes A2 and B2. The report baseline scores 0 on A2, B2 and D1 for every model, while the two GPT models reach 82.1 and 86.9 on D1 through iteration.

\begin{table}[htbp]
  \centering
  \caption{Factor-pool scores by task (0--100). The report-baseline row is the mean over the 4 models, and the last column counts passes. The first letter of a task ID is its difficulty tier at selection (Table~\ref{tab:tasks}).}
  \label{tab:per-task}
  \small
  \setlength{\tabcolsep}{4.5pt}
  \begin{tabular}{@{}lcccccccc@{}}
    \toprule
    & A1 & A2 & B1 & B2 & C1 & C2 & D1 & Pass \\
    \midrule
    gpt-6.1-sol & 39.5 & 70.2 & 89.6 & 66.7 & 94.3 & 80.8 & 82.1 & 7/7 \\
    gpt-6.1-terra & 74.5 & 0 & 92.2 & 0 & 62.9 & 79.9 & 86.9 & 5/7 \\
    deepseek-v4.1-flash & 56.0 & 0 & 76.0 & 0 & 53.8 & 65.0 & 0 & 4/7 \\
    muse-spark-1.3 & 41.8 & 0 & 61.9 & 0 & 0 & 79.7 & 0 & 3/7 \\
    Report baseline & 21.7 & 0 & 51.5 & 0 & 42.6 & 61.7 & 0 & 12/28 \\
    \bottomrule
  \end{tabular}
\end{table}

\begin{table}[htbp]
  \centering
  \caption{Pairwise comparisons. Upper part: differences in factor-pool score between models (first minus second) with the $p$-values of two tests. Lower part: differences between the factor pool and each control (four models pooled), with 95\% intervals; ``Pool higher'' and the Wilcoxon $p$ count only runs where neither set is empty. Test methods are in Appendix~\ref{app:scoring}.}
  \label{tab:pairwise}
  \small
  \begin{tabular}{@{}lccc@{}}
    \toprule
    Model pair & Score difference & Trading-day $p$ & Task $p$ \\
    \midrule
    gpt-6.1-sol $-$ gpt-6.1-terra & 18.1 & 0.012 & 0.359 \\
    gpt-6.1-sol $-$ deepseek-v4.1-flash & 38.9 & $<$0.001 & 0.062 \\
    gpt-6.1-sol $-$ muse-spark-1.3 & 48.5 & $<$0.001 & 0.047 \\
    gpt-6.1-terra $-$ deepseek-v4.1-flash & 20.8 & $<$0.001 & 0.062 \\
    gpt-6.1-terra $-$ muse-spark-1.3 & 30.4 & 0.001 & 0.062 \\
    deepseek-v4.1-flash $-$ muse-spark-1.3 & 9.6 & 0.113 & 0.500 \\
    \midrule
    Control & Difference interval & Pool higher & Wilcoxon $p$ \\
    \midrule
    Pool $-$ report baseline & [16.9, 31.6] & 12/14 & 0.004 \\
    Pool $-$ own pick & [5.7, 15.9] & 19/25 & 0.005 \\
    Pool $-$ best in training & [2.7, 12.6] & 15/25 & 0.013 \\
    Pool $-$ training top-20 mean & -- & 21/25 & $<$0.001 \\
    \bottomrule
  \end{tabular}
\end{table}

\subsection{Where the Gap Appears}
\label{sec:stage}
The four models start from similar points, and the gap forms during iteration. Their report baselines are 31.0, 17.0, 28.1 and 25.5 points (in the order sol, terra, deepseek, muse, likewise below), showing no advantage for the two GPT models; after 6 rounds of iteration the pools improve on the report baselines by 43.7, 39.6, 7.7 and 0.7 points (Figure~\ref{fig:headline}, right).

There is no difference on the training period; the models separate only on the test period. On the training period, the mean $S$ of the pools' factors is 3.86, 3.74, 3.86 and 3.68, and the mean of the training-period top 20 is 3.93, 3.83, 3.73 and 3.67, overlapping each other (Appendix Figure~\ref{fig:layers-train}).
On the test period, the top-20 mean $S$ spreads to 3.24, 2.78, 2.29 and 2.03, and the $S$ of non-empty pools is 4.48, 4.28, 2.87 and 2.30.
In other words, the four models' candidates look equally good on the training period but perform differently on new data. Section~\ref{sec:analysis} analyzes which research behaviors this divergence is associated with.

\subsection{Efficiency}
\label{sec:efficiency}
Figure~\ref{fig:tokens} plots scores against token usage; details of the research process are in Appendix Table~\ref{tab:longhorizon}.
gpt-6.1-terra uses about a third of the time and half of the tokens of gpt-6.1-sol and scores second only to it; deepseek-v4.1-flash uses about 2.7 times the tokens of gpt-6.1-sol yet scores less than half as much. Token usage does not move with the score: spending more tokens does not mean doing better research.
Scoring ignores trading costs; a backtest simulation on a long-only, after-cost basis is in Appendix~\ref{app:economics}.

\begin{figure}[htbp]
  \centering
  \includegraphics[width=0.6\linewidth]{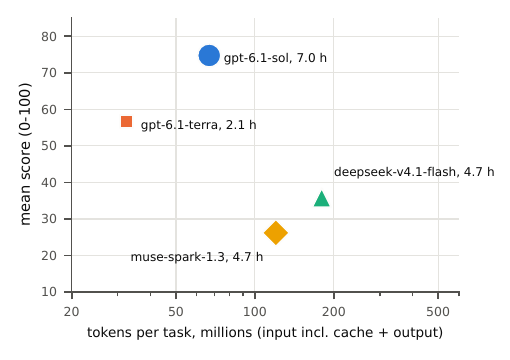}
  \caption{Average factor-pool score against tokens per task (log scale, input including cache hits plus output). Larger markers mean longer average time; the time next to each marker is the mean over the 7 tasks, whereas Appendix Table~\ref{tab:longhorizon} reports the median.}
  \label{fig:tokens}
\end{figure}

\section{Analysis}
\label{sec:analysis}
\begin{figure}[htbp]
  \centering
  \includegraphics[width=0.62\linewidth]{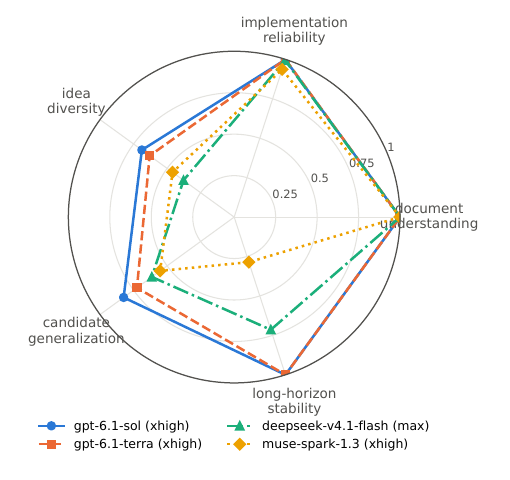}
  \caption{Normalized comparison of the five abilities (further out is better; for relative comparison between models only). Document understanding is the plan's recall of core methods (complete in all 28 runs; the axis shows the absence of a difference); implementation reliability is 1 minus the share of hypotheses failing the engineering checks; idea diversity is 1 minus the same-round correlation; candidate generalization is the ratio of the model-level mean test-period $S$ to the model-level mean training-period $S$ of the training top-20 candidates (a property of the candidates as a whole, not the pool's train--test gap in Table~\ref{tab:ability}); long-horizon stability is the share of runs without a requested rewrite, record drift or debugging loop. 7 runs per model.}
  \label{fig:radar}
\end{figure}

Section~\ref{sec:stage} showed that the gap forms during iteration and shows up only on the test period. This section analyzes which research behaviors it is associated with.
Table~\ref{tab:ability} summarizes the process metrics by ability, and Figure~\ref{fig:radar} plots them together after normalization.
Document understanding and code writing can be ruled out first: the research plans of all 28 runs cover all of the report's core methods and the report baselines are similar; apart from muse-spark-1.3, each model has at most 0.6\% of hypotheses failing the engineering checks, while deepseek-v4.1-flash, with almost no engineering problems, also ranks in the bottom two.
The largest differences are in hypothesis and iteration and in generalization, where the GPT models are clearly higher on the idea-diversity and candidate-generalization axes of Figure~\ref{fig:radar}. Below we first walk through one complete research run with a pair of cases, then examine these two abilities, and finally discuss engineering and long-horizon stability.
Each case is a single run and only illustrates a mechanism; it does not constitute causal evidence. Cases are written as ``task-model'', e.g.\ D1-sol.

\begin{table}[htbp]
  \centering
  \caption{Process metrics by ability (7 runs per model; definitions in Table~\ref{tab:process}). The total is the factor-pool score with its 95\% interval in parentheses; the train--test gap counts non-empty pools only; tokens are input (including cache hits) plus output.}
  \label{tab:ability}
  \footnotesize
  \setlength{\tabcolsep}{3.5pt}
  \resizebox{\linewidth}{!}{\begin{tabular}{@{}lccccccccc@{}}
    \toprule
    & Total & Implementation & \multicolumn{3}{c}{Hypothesis and iteration} & \multicolumn{2}{c}{Generalization} & Long horizon & Tokens per task \\
    \cmidrule(lr){4-6}\cmidrule(lr){7-8}
    Model & (95\% interval) & failed checks & same-round corr. & improvement & factors in pool & train--test gap & overfitting & rewrite/drift & (millions) \\
    \midrule
    gpt-6.1-sol & 74.7 (56.6--86.3) & 0.2\% & 0.31 & 0.36 & 5.7 & $-0.62$ & 0 & 0/0 & 67 \\
    gpt-6.1-terra & 56.6 (45.3--72.9) & 0.0\% & 0.37 & 0.34 & 3.7 & $-0.54$ & 0 & 0/0 & 32 \\
    deepseek-v4.1-flash & 35.8 (21.5--52.9) & 0.6\% & 0.62 & 0.41 & 3.1 & 0.99 & 2 & 0/0 & 180 \\
    muse-spark-1.3 & 26.2 (15.3--41.4) & 6.4\% & 0.54 & 0.47 & 2.6 & 1.38 & 3 & 4/1 & 120 \\
    \bottomrule
  \end{tabular}}
\end{table}

\subsection{A Complete Case: D1-sol and D1-deepseek}
\label{sec:case}
\begin{figure}[htbp]
  \centering
  \includegraphics[width=\linewidth]{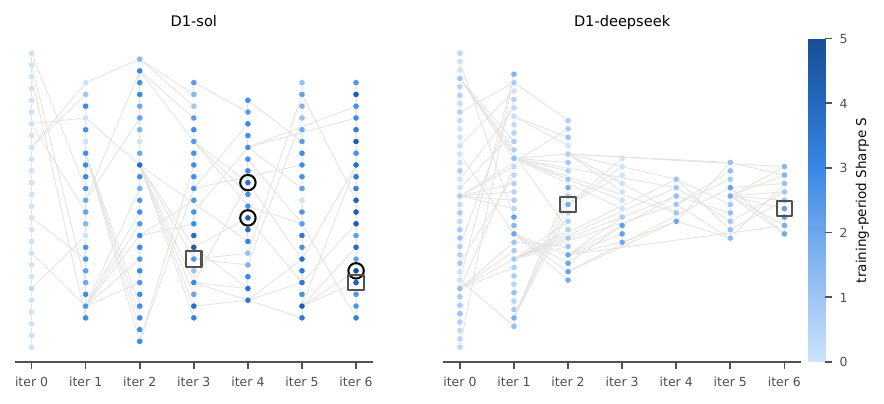}
  \caption{Iteration trees of D1-sol (left) and D1-deepseek (right). Each dot is a candidate that passed the engineering checks; the horizontal axis is the round, lines are parent--child links and color is training-period $S$; black rings mark the pool's factors and gray squares the factors the model picked itself. D1-deepseek's pool is empty.}
  \label{fig:lineage}
\end{figure}

We chose this pair because they differ the most on the same task: D1-sol's factor pool scores 82.1, while D1-deepseek's pool is empty. Figure~\ref{fig:lineage} shows their iteration trees side by side.
\begin{enumerate}
  \item Reading the report. Both found that the report omits key parameters, and both stated in their plans how they would fill the gaps. D1-sol treated the report as a chain of necessary conditions to be tested, broken into 7 components and 8 research directions; D1-deepseek treated it as a model that only needs tuning, and mainly ran grid searches over window length, prediction horizon and similarity function.
  \item Getting started. In round 0 D1-sol proposed 26 hypotheses; besides the report formula, the same round already included controls that ``remove just one component of the original''. D1-deepseek proposed 36, with controls mostly drawn from a parameter grid. For both, round-0 feedback showed that the report's direction does not hold on the training period, and both report baselines are empty pools.
  \item Iterating. D1-sol used its controls to reject the original pattern explanation: removing some components left the results unchanged, so they were not necessary conditions. It then switched to an intraday-data direction proposed separately in round 0, and every later change of direction was triggered by a ``change one component only'' control. D1-deepseek kept the strongest training-period branch; after the controls showed limited gains, the object of study stayed the same, and later changes concentrated on parameters, weights and combination.
  \item Converging. D1-sol kept 18--26 hypotheses per round, and the share of controls rose from 5/26 in round 0 to 14/21 in round 6; D1-deepseek fell from 36 to 6--11, and its last two rounds were mostly combinations of existing branches.
  \item Outcome. In the end 3 low-correlation factors entered D1-sol's pool, with test-period $S=4.92$; the highest training-period $S$ among all 123 candidates of D1-deepseek was only 2.29, none reached the entry threshold of 3, and its pool is empty.
\end{enumerate}
This pair corresponds to the two differences discussed below: whether hypotheses are diverse (\S\ref{sec:hypothesis}), and whether good training-period results carry over to the test period (\S\ref{sec:generalization}).

\subsection{Hypothesis and Iteration: More Diverse Ideas per Round, Higher Pool Scores}
\label{sec:hypothesis}
\begin{figure}[htbp]
  \centering
  \includegraphics[width=0.6\linewidth]{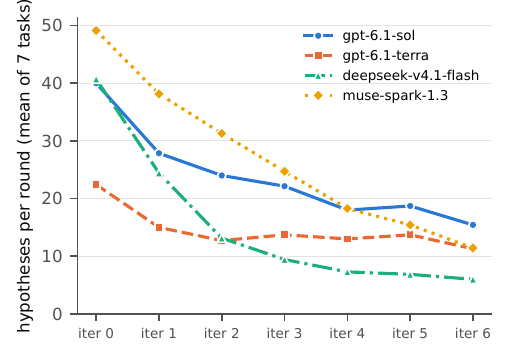}
  \caption{Hypotheses proposed per round (mean over 7 tasks).}
  \label{fig:rounds}
\end{figure}

The mean correlation among candidates of the same round is 0.31 and 0.37 for the two GPT models, against 0.62 and 0.54 for the other two; across the 28 runs, its rank correlation with the pool score is $-0.43$ ($p=0.024$), that is, the less similar the ideas within a round, the higher the pool score.
Figure~\ref{fig:rounds} shows how the ideas thin out round by round: deepseek-v4.1-flash has only 6.0 hypotheses on average left in round 6, while gpt-6.1-sol still keeps about 15. In later rounds the latter two models focus more and more on local tweaks; their training-period results improve more easily, but the candidates within a round become more and more alike. Besides D1 (\S\ref{sec:case}), C1 shows the same pattern: after round 0, C1-sol still had 21--29 low-correlation hypotheses per round together with systematic controls, and 9 factors entered its pool for 94.3 points; from round 2 on, C1-deepseek had only 2--4 per round, almost all tweaks to the combination weights of three factors, and 4 factors entered its pool for 53.8 points. Diverse controls can also simplify a factor: B1-sol used controls that removed conditions one at a time to show that the ``volume surge'' condition stressed by the report can be dropped.
Diverse hypotheses directly add effective factors that pass the correlation limit: the two GPT models' pools hold 5.7 and 3.7 factors on average, against 3.1 and 2.6 for the other two models; across the 28 runs, the rank correlation between the number of factors in the pool and the score is 0.68.

\subsection{Generalization: Do Good Training-Period Results Carry Over to the Test Period?}
\label{sec:generalization}
The training-period $S$ of the four models' pools is almost the same (3.68--3.86), but the pools of deepseek-v4.1-flash and muse-spark-1.3 fall back by 0.99 and 1.38 respectively on the test period, with 5 runs judged as training-period overfitting by rule; the two GPT models have none, and their train--test gaps are even negative ($-0.62$ and $-0.54$), because combining several low-correlated factors raises the Sharpe ratio. Tasks A2 and B2 are the clearest examples: deepseek-v4.1-flash and muse-spark-1.3 both have factors that enter the pool with training-period $S\ge3$, yet their test-period $S$ is only 0.99--1.35 and none passes, while gpt-6.1-sol's pools on these two tasks reach a test-period $S$ of 4.21 and 4.00. The entry threshold is not lax either: no candidate of D1-deepseek or B2-terra reaches a training-period $S$ of 3, so their pools are empty.
Since the pool rule looks only at training-period results, this part of the gap reflects whether the candidates themselves hold up on new data, not how well they were picked.

\subsection{Engineering Reliability and Long-Horizon Stability}
\label{sec:reliability}
Engineering problems concentrate in muse-spark-1.3: 6.4\% of its hypotheses fail the engineering checks, it was asked to rewrite its output 4 times, and its research records drifted once. Judged by the written rules (Appendix~\ref{app:annotation}), debugging loops (3 or more engineering check failures and unchanged-code resubmissions in total) occur for deepseek-v4.1-flash (2 runs) and muse-spark-1.3 (3 runs); report misreading and wrong use of data fields never occur (run-by-run results in Appendix Table~\ref{tab:fullresults}). The most serious case is C1-muse: in the coding stage it applied a filter that should only act on previous-day data to same-day data, so every candidate uses look-ahead, and all 10 factors that would have entered the pool are removed by the look-ahead check, leaving an empty pool. Context compaction by itself does not decide success: gpt-6.1-sol's conversation history was compacted 28 times without failures or drift in its records, and after one compaction B2-muse's summary mixed up data fields, which the model later corrected itself, so the final output did not drift.
Engineering failures are therefore concentrated in a few runs and cannot explain the overall gap between models.

\section{Limitations}
\label{sec:limitations}
\begin{enumerate}
  \item Small sample. The task set has only 7 tasks and each model runs once per task, so the randomness of a single run is not estimated; the model ranking should be read as an observation from these runs, not as a stable ranking. We plan to add tasks from the 431 reports that run through the pipeline and to repeat runs for each model to estimate run-to-run variation.
  \item Strong prompt guidance. We evaluate models under fairly strong research-method prompts (\S\ref{sec:prompts}), so scores reflect the model and the prompts together, and how much the prompts reduce overfitting is supported only by researchers' informal observations. We plan to remove the research-method guidance, keep only necessary instructions such as the data interface, time boundaries and delivery format, let the model decide how to iterate and pick, and compare with the results here.
  \item Pre-training may have seen the answers. The evaluated models may have seen the original reports and factor-tracking reports covering 2024--2025 during pre-training \citep{liu2026finforesight}, so test-period scores may overstate true predictive ability; a test window later than all models' training cutoffs could be added.
  \item Parts of the pool rule are untested. The entry threshold of 3 and the correlation threshold of 0.6 are empirical and have not been subjected to sensitivity analysis; the training backtests on which pool selection relies are stored in a directory the model can write to, and were not rerun at scoring time; look-ahead was checked only by log search and static source analysis, without day-by-day perturbation tests, so the result should be read as ``no evidence of look-ahead found''.
  \item Annotations not yet reviewed by humans. Plan recall, failure types and other annotations were judged by a single LLM annotator following written rules and should be read as preliminary.
  \item Limited scope. The conclusions are limited to our research workflow, the Codex CLI toolchain and A-share price--volume and order-flow reports, and cannot be extrapolated directly to fully autonomous research or other kinds of research.
\end{enumerate}

\section{Conclusion}
\label{sec:conclusion}
We built an end-to-end quantitative research task from real brokerage reports: the report is only a starting point, and over 7 rounds of iteration the model proposes hypotheses, writes factors and improves them from feedback; after the research, a fixed rule that looks only at training-period results selects a low-correlated factor pool, which is scored on unseen market data after removing styles.
The evaluation of 4 models yields three conclusions. First, most factor pools pass on the test period, with an average score clearly above the report baseline and above the submissions the models picked themselves. Second, the four models' report baselines are similar and their candidates look equally good on the training period; the gap forms during iteration and shows up only on the test period. Third, the leading models propose more diverse ideas per round, keep more effective factors in the pool and fall back less on the test period; neither reading the report nor writing code is a major source of the gap.
Evaluating research agents therefore cannot rely only on how well they do on known data: scoring should be done on frozen future data after removing styles, and ``what the research produced'' should be measured separately from ``what was finally picked''.
Combining a repeatable research loop with objective scoring on future data may also offer a way to evaluate agents on other noisy decisions with interacting factors, but that transferability remains to be tested in other domains.
Next, we will enlarge the task set and repeat runs, and remove the research-method prompts while keeping only necessary guidance, to test whether models researching autonomously still avoid in-sample overfitting.

\bibliographystyle{plainnat}
\renewcommand{\bibfont}{\footnotesize}
\bibliography{main}

\appendix
\section{Glossary}
\label{app:glossary}
Table~\ref{tab:glossary} summarizes the financial terms used in the main text and the names defined in this paper.

\begin{table}[H]
  \centering
  \caption{Terms used in the main text.}
  \label{tab:glossary}
  \small
  \begin{tabular}{@{}p{3.6cm}p{11.0cm}@{}}
    \toprule
    English & Meaning \\
    \midrule
    factor / alpha & A score given to each stock every day to predict its future relative return \\
    cross-section & The set of all stocks at the same point in time \\
    quantile layer & Stocks are split into 10 equal groups by factor value, and the returns of the groups are compared \\
    style-free return & The return remaining after removing the part explained by the market, industries and styles \\
    style neutralization & Removing the association between a factor and styles such as size, value and momentum \\
    Sharpe ratio & Mean return divided by return volatility, then annualized \\
    look-ahead bias & Using information that could only be known after the decision time \\
    turnover & Daily traded amount as a share of capital \\
    training / test period & 2022--2023, whose backtest results the model can see during research; 2024--2025, used only for scoring \\
    candidate & A factor that has passed the engineering checks and the training-period backtest \\
    ablation & A control experiment that removes or replaces only one component and keeps the rest unchanged \\
    factor pool & A set of low-correlation factors selected from all candidates by a fixed rule using training-period results only; the object of the main score \\
    report baseline & The factor pool obtained by applying the same rule to round-0 candidates only \\
    pass & Test-period $S\ge2$ \\
    overfitting & Factors picked after repeated trial and error on the training period look good only on the training period and fail on new data \\
    \bottomrule
  \end{tabular}
\end{table}

\section{Scoring Method}
\label{app:scoring}

\subsection{Time split}
The training replay runs from 2022-01-04 to 2023-12-29; the first 60 trading days are used only for warm-up, and training-period performance is counted from 2022-04-07. The test replay starts on 2023-10-09; the first 60 trading days are used only to warm up the combination weights, and the scoring period is 2024-01-02 to 2025-12-31, 485 trading days in total.

\subsection{Factor pool}
Training-period $S$ is computed with Eq.~\eqref{eq:sharpe} on the training scoring window (2022-04-07 to 2023-12-29); candidates whose training statistics are undefined (e.g.\ zero volatility) are not ranked. Correlation uses the long-only style-free daily return output by the training backtest (the position build-up day removed), computed as a Pearson correlation over the trading days the two candidates share; the 0.6 threshold is not applied to the absolute value. The look-ahead check is run after the pool is built, as described in Appendix~\ref{app:annotation}; factors judged as ``look-ahead'' or ``not traceable'' are both removed. The rule is implemented as \texttt{select\_pool} in the scoring module and is run by an independent container after the research ends.

\subsection{Single-factor processing}
For each selected factor, only the value on the opening cross-section of each day is kept; untradable stocks are removed (suspended, closed locked at the upper or lower price limit on either of the previous two days, or outside the universe), and a cross-sectional nonlinear neutralization is applied against 6 styles (size, value, momentum, beta, residual volatility and liquidity). Training feedback and scoring use the same processing.

\subsection{Multi-factor combination}
Each factor receives a combination weight from its own daily performance over the previous 60 trading days; only factors with positive weight enter the combination, and the combined signal $C_{i,d}$ is their weighted average. If no factor has a positive weight, the one with the highest weight is used. The combined signal goes through the tradability filter and style neutralization once more. The scorer never flips a factor's direction.

\subsection{Layers and style-free returns}
At each open, tradable stocks are split into 10 equal layers by $C_{i,d}$ (L0 has the highest signal); each layer is an equal-weight long-only account rebalanced at the volume-weighted average price of the first 30 minutes after the open, with no transaction costs. Style-free returns are computed once for the whole market rather than by a separate regression for each layer: every day the returns of all stocks are regressed across stocks on 10 Barra CNE5 styles, 32 industries and the market \citep{orr2012cne5},
\begin{equation}
  r_i=c+\sum_{k=1}^{10}b_k X_{k,i}+\sum_{j=1}^{32}g_j I_{j,i}+e_i ,
  \label{eq:spec}
\end{equation}
giving each stock's residual return $e_i$; settling each layer's holdings on these residuals gives the layer's style-free return $r_{k,d}$.

\subsection{Score}
The daily excess return is the mean of the top five layers minus the ten-layer mean, which equals half of the spread between the top and bottom halves:
\begin{equation}
  x_d=\frac15\sum_{k=0}^{4}r_{k,d}-\frac1{10}\sum_{k=0}^{9}r_{k,d},
  \qquad
  S=\frac{\bar{x}}{s_x}\sqrt{250},
  \label{eq:sharpe}
\end{equation}
where $s_x$ is the sample standard deviation (with one degree of freedom removed). The score of a run is
\begin{equation}
  \mathrm{score}=\begin{cases}0, & S<2,\\ \min(S/6,\,1), & S\ge 2,\end{cases}
  \label{eq:score}
\end{equation}
and an empty pool scores 0; if the replay has missing trading days, non-finite values or zero standard deviation, no score is produced and the case is recorded separately as an infrastructure error. Under the approximation of i.i.d.\ returns, $t\approx S\sqrt{T/250}$ \citep{lo2002sharpe}; with $T=485$, $S=2$ corresponds to about $t\approx2.79$.

\subsection{Intervals and tests}
The 485 trading days of the test period are resampled 2,000 times with 20-day circular blocks (random seed 0), and the same set of indices is used for all runs and all controls; runs with an empty pool, an empty set or no model-selected submission always score 0. Intervals for average scores take the 2.5\% and 97.5\% quantiles; for the ``pool minus control'' difference, both are computed on each resample before taking quantiles. Between models we compare by trading day to obtain two-sided $p$-values, and by task we apply an exact sign-flip test to the score differences on the 7 tasks ($2^7$ combinations). Run-by-run paired comparisons between the pool and the controls use test-period $S$ (scores are both 0 whenever $S<2$, which produces too many ties), with the Wilcoxon signed-rank test.

\section{Task Set Selection}
\label{app:taskset}
The task set was selected from historical runs using training-period information only. Only reports published on or before 31 December 2023 were eligible; the candidate pool consisted of 208 historical runs meeting the date rule, and after removing runs with outdated artifact formats, runs that did not complete all 7 rounds, runs whose titles were weekly or factor-tracking reports, and runs that took too long, 87 reports qualified.
On the training period we computed an approximate score for each historical run (called the proxy score, following the old protocol used at selection time: the Sharpe ratio of the long-only style-free return, divided by twice the square root of mean daily one-way turnover). Let $b_0$ be the best score in round 0 and $b_L$ the best score in rounds 1--6; four tiers follow: A improvable by research ($b_L-b_0\ge3$; 14 eligible, 2 selected), B strong baseline ($b_0\ge7$ and $b_L\le b_0$; 3 eligible, 2 selected), C high ceiling ($b_0\ge7$ and $b_L>b_0$; 9 eligible, 2 selected) and D hard ($b_L<4.5$; 1 eligible, 1 selected). Within each tier we deduplicated by topic and data type, and ensured that every year from 2020 to 2023 is represented.
This process has three known biases: the proxy scores follow the old protocol; the iteration feedback of the historical runs included 2024 market data, so $b_L$ may be optimistic; and reports published in 2022--2023 overlap the training period, but the entire test period postdates the publication dates of all reports.

\begin{table}[H]
  \centering
  \caption{Original titles and selection proxy scores of the 7 reports in the task set (old protocol, used only for selection and not comparable with our scores).}
  \label{tab:titles}
  \small
  \begin{tabular}{@{}llp{9.4cm}r@{}}
    \toprule
    ID & Broker & Title (translated from Chinese) & $b_0/b_L$ \\
    \midrule
    A1 & Founder Securities & Multi-factor stock selection series no.~4: identifying stock-level momentum and constructing the ``team coin'' factor & $-1.64/7.63$ \\
    A2 & Zheshang Securities & Building stock selection strategies under price jumps: information shocks and selecting ``key'' targets & $1.44/6.52$ \\
    B1 & Founder Securities & Multi-factor stock selection series no.~1: alpha information in moments of surging volume & $9.30/9.27$ \\
    B2 & Huaan Securities & Quantitative fundamentals series no.~10: attraction and repulsion between stock price and money flow & $7.20/7.20$ \\
    C1 & Kaiyuan Securities & Market microstructure series no.~15, high-frequency factors: characterizing large-player behavior from the per-minute average trade size & $10.10/13.14$ \\
    C2 & Soochow Securities & ``Technical analysis meets stock-selection factors'' series no.~5: the shifted CPV factor and stock-selection information in price-volume autocorrelation & $9.06/11.05$ \\
    D1 & Dongguan Securities & Quantitative special report: how much do similar candlesticks tell? & $2.60/4.43$ \\
    \bottomrule
  \end{tabular}
\end{table}

\section{Process Metrics and Annotation Rules}
\label{app:annotation}
\subsection{Automatic metrics}
All are computed from the run artifacts.
The training-period $S$ of candidates is computed with the scoring formula on the training period (2022-04-07 to 2023-12-29) for every candidate that passed the engineering checks, 3,897 in total;
the check-failure rate is $1-$ number of candidates passing the checks $/$ number of proposed hypotheses, and the counts of engineering check failures, unchanged source, rewrite requests, retries and timeouts are taken from the research logs;
within-round correlation is the mean pairwise Pearson correlation of training-period daily returns among candidates of the same round, then averaged over rounds;
the improvement rate is, among child factors whose parents all exist, the proportion whose training-period $S$ is higher than that of all their parents;
the train--test gap is the mean training-period $S$ of the pool's factors minus the pool's test-period $S$ (empty pools excluded);
context compaction is taken from compaction events in the session records.

\subsection{Annotation rules}
Annotation is done by one LLM annotator following written rules; every judgment is accompanied by a location in the source or a short quote, pending human review.
\begin{itemize}
  \item Formula completeness (per report): complete / missing parameters / idea only; the 2--6 core methods that the report presents as its main contribution are also listed;
  \item Plan recall (per run): number of core methods that the plan explicitly mentions with a location in the report $/$ total number of core methods;
  \item Failure types: report misreading, field-definition error, look-ahead, direction or scale error, and context drift are judged by the annotator; training-period overfitting (train--test gap $\ge1.5$ and pool test-period $S<2$), debugging loop (engineering check failures plus unchanged source totaling at least 3) and empty pool are determined directly from the numbers.
\end{itemize}
The review removed the ``direction or scale error'' label from 10 runs that ``found negative monotonicity in round 0 and reversed it later'', since this is normal research behavior.

\subsection{Log and source checks}
We searched every command executed in the 28 runs and its output, and found 107 records related to test-period dates or the scoring configuration. After reviewing each of them, all are time-boundary statements in the prompts and plans, dates in header-file comments, or the single read of the scoring configuration; there is no command reading test-period market data.
The look-ahead check covers all 3,897 candidates: it first locates places that read, by the current-day index, fields not yet known at the open (prices, trades, order book and so on, excluding the universe, suspension and theoretical price limits), and then traces along the function call chain whether that index equals the current day. Of the 627 flagged candidates, 417 have indices that trace back to the previous trading day in every case; the remaining 210 all come from C1-muse and read the day's close in the tradability check, consistent with the conclusion of the model's own check in the selection stage. The static check cannot detect aliasing or pointer arithmetic, so it serves only as a necessary condition.

\section{Full Results}
\label{app:fullresults}
Model names in the tables are short names: sol, terra, deepseek and muse correspond to gpt-6.1-sol, gpt-6.1-terra, deepseek-v4.1-flash and muse-spark-1.3, respectively.

\subsection{Per-Run Results}
Table~\ref{tab:fullresults} combines the factor pools, controls and annotations of the 28 runs. Under ``factors'', the two columns are the numbers of factors in the factor pool and in the report baseline; the two $S$ columns are Sharpe ratios, where the training $S$ is the mean training-period $S$ of the pool's factors and the test $S$ is that of the combined pool; the five columns under ``score'' are test-period scores on a 0--100 scale for the factor pool, the report baseline, the model's own pick, the best in training and the training top-20 mean. Failure-type abbreviations: ov = training-period overfitting, lp = debugging loop, em = empty pool, la = look-ahead, dr = context drift, dir = direction or scale error.
{\scriptsize\setlength{\tabcolsep}{3.5pt}
\begin{longtable}{@{}llrrrrrrrrrrl@{}}
  \caption{Per-run results of the 28 runs.}\label{tab:fullresults}\\
  \toprule
  & & \multicolumn{2}{c}{Factors} & \multicolumn{2}{c}{Sharpe $S$} & \multicolumn{5}{c}{Score (0--100)} & & \\
  \cmidrule(lr){3-4}\cmidrule(lr){5-6}\cmidrule(lr){7-11}
  Task & Model & Pool & Base & Train & Test & Pool & Base & Own & Best & Top-20 & Hours & Failure types \\
  \midrule
  \endfirsthead
  \toprule
  & & \multicolumn{2}{c}{Factors} & \multicolumn{2}{c}{Sharpe $S$} & \multicolumn{5}{c}{Score (0--100)} & & \\
  \cmidrule(lr){3-4}\cmidrule(lr){5-6}\cmidrule(lr){7-11}
  Task & Model & Pool & Base & Train & Test & Pool & Base & Own & Best & Top-20 & Hours & Failure types \\
  \midrule
  \endhead
  A1 & sol & 6 & 3 & 3.58 & 2.37 & 39.5 & 0.0 & 37.1 & 46.7 & 25.0 & 6.8 & -- \\
  A1 & terra & 5 & 1 & 3.94 & 4.47 & 74.5 & 0.0 & 56.5 & 55.5 & 59.5 & 2.7 & -- \\
  A1 & deepseek & 5 & 1 & 3.64 & 3.36 & 56.0 & 35.4 & 38.9 & 38.9 & 40.9 & 4.8 & -- \\
  A1 & muse & 2 & 2 & 3.68 & 2.51 & 41.8 & 51.5 & 0.0 & 0.0 & 7.4 & 5.5 & dr, lp \\
  A2 & sol & 3 & 0 & 3.74 & 4.21 & 70.2 & 0.0 & 0.0 & 58.8 & 55.6 & 5.8 & -- \\
  A2 & terra & 2 & 0 & 3.15 & 1.93 & 0.0 & 0.0 & 0.0 & 0.0 & 9.7 & 1.6 & -- \\
  A2 & deepseek & 3 & 0 & 3.40 & 1.21 & 0.0 & 0.0 & 0.0 & 0.0 & 0.0 & 3.6 & ov, lp \\
  A2 & muse & 1 & 0 & 3.44 & 1.17 & 0.0 & 0.0 & 0.0 & 0.0 & 0.0 & 2.9 & ov \\
  B1 & sol & 6 & 2 & 3.45 & 5.38 & 89.6 & 70.6 & 78.1 & 82.9 & 58.0 & 6.9 & -- \\
  B1 & terra & 4 & 0 & 3.56 & 5.53 & 92.2 & 0.0 & 69.9 & 77.5 & 64.1 & 2.0 & -- \\
  B1 & deepseek & 3 & 2 & 4.43 & 4.56 & 76.0 & 61.9 & 68.7 & 61.7 & 65.6 & 6.0 & lp \\
  B1 & muse & 1 & 1 & 4.04 & 3.71 & 61.9 & 73.6 & 77.6 & 61.9 & 64.8 & 3.8 & -- \\
  B2 & sol & 2 & 0 & 3.57 & 4.00 & 66.7 & 0.0 & 0.0 & 57.2 & 39.9 & 7.0 & -- \\
  B2 & terra & 0 & 0 & -- & -- & 0.0 & 0.0 & 0.0 & 0.0 & 1.7 & 2.2 & em \\
  B2 & deepseek & 1 & 0 & 3.45 & 0.99 & 0.0 & 0.0 & 0.0 & 0.0 & 0.0 & 5.9 & ov \\
  B2 & muse & 3 & 0 & 3.37 & 1.35 & 0.0 & 0.0 & 0.0 & 0.0 & 0.0 & 7.9 & ov, lp \\
  C1 & sol & 9 & 2 & 3.88 & 5.66 & 94.3 & 74.9 & 64.3 & 66.5 & 60.5 & 7.8 & -- \\
  C1 & terra & 5 & 2 & 3.40 & 3.78 & 62.9 & 51.3 & 45.0 & 44.8 & 34.1 & 2.0 & -- \\
  C1 & deepseek & 4 & 2 & 4.06 & 3.23 & 53.8 & 44.2 & 0.0 & 66.6 & 41.7 & 3.7 & -- \\
  C1 & muse & 0 & 0 & -- & -- & 0.0 & 0.0 & 0.0 & 0.0 & 0.0 & 4.8 & la, em \\
  C2 & sol & 11 & 6 & 4.24 & 4.85 & 80.8 & 71.3 & 77.8 & 54.5 & 49.8 & 7.7 & -- \\
  C2 & terra & 7 & 3 & 4.13 & 4.79 & 79.9 & 67.3 & 69.8 & 42.1 & 45.4 & 2.1 & -- \\
  C2 & deepseek & 6 & 4 & 4.20 & 3.90 & 65.0 & 55.2 & 60.6 & 60.6 & 57.9 & 3.3 & -- \\
  C2 & muse & 10 & 4 & 4.27 & 4.78 & 79.7 & 53.1 & 84.4 & 86.7 & 65.6 & 4.8 & -- \\
  D1 & sol & 3 & 0 & 4.59 & 4.92 & 82.1 & 0.0 & 95.9 & 83.6 & 74.2 & 6.7 & -- \\
  D1 & terra & 3 & 0 & 4.25 & 5.21 & 86.9 & 0.0 & 79.1 & 100.0 & 66.2 & 2.0 & -- \\
  D1 & deepseek & 0 & 0 & -- & -- & 0.0 & 0.0 & 0.0 & 0.0 & 0.0 & 5.5 & em \\
  D1 & muse & 1 & 0 & 3.27 & 0.29 & 0.0 & 0.0 & 0.0 & 0.0 & 0.0 & 3.2 & dir, ov, lp \\
  \bottomrule
\end{longtable}
}

\subsection{Economic Meaning of the Research Results}
\label{app:economics}
Scoring looks only at style-free stock-selection ability. To give an intuitive sense of magnitude, we put the long-only portfolios of each model's 7 factor pools together with equal weights and compute the 2024--2025 excess return over the index of the tradable universe (tasks with an empty pool hold cash; short selling is restricted in A-shares, so only the long side is considered). Before costs, the four models' annualized excess returns are 6.6\%--10.1\% (information ratios 3.13--5.20); mean daily one-way turnover is 11\%--19\%, and after a one-way cost of 0.1\% they become $-0.5$\%--1.7\% (information ratios $-0.27$--0.97). Long-only excess returns still contain style returns, so the ranking does not fully match the scores. These are backtest simulations, not live trading, and do not constitute investment advice.

\subsection{Selection Methods in the Training Period}
\begin{figure}[H]
  \centering
  \includegraphics[width=0.42\linewidth]{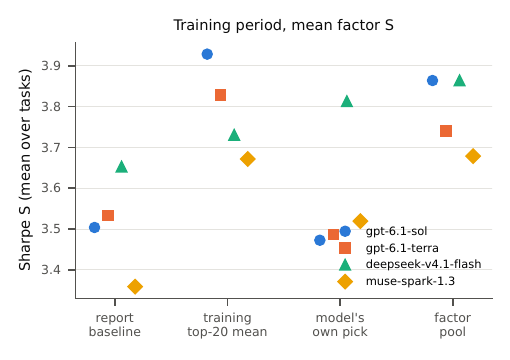}
  \caption{Average training-period Sharpe ratio $S$ of the factors chosen by four selection methods (mean of the standalone training-period $S$ of each set's factors, then averaged over the 7 tasks; empty pools excluded).}
  \label{fig:layers-train}
\end{figure}

\subsection{Cumulative Test-Period Returns}
\label{app:cumulative}
\begin{figure}[H]
  \centering
  \includegraphics[width=\linewidth]{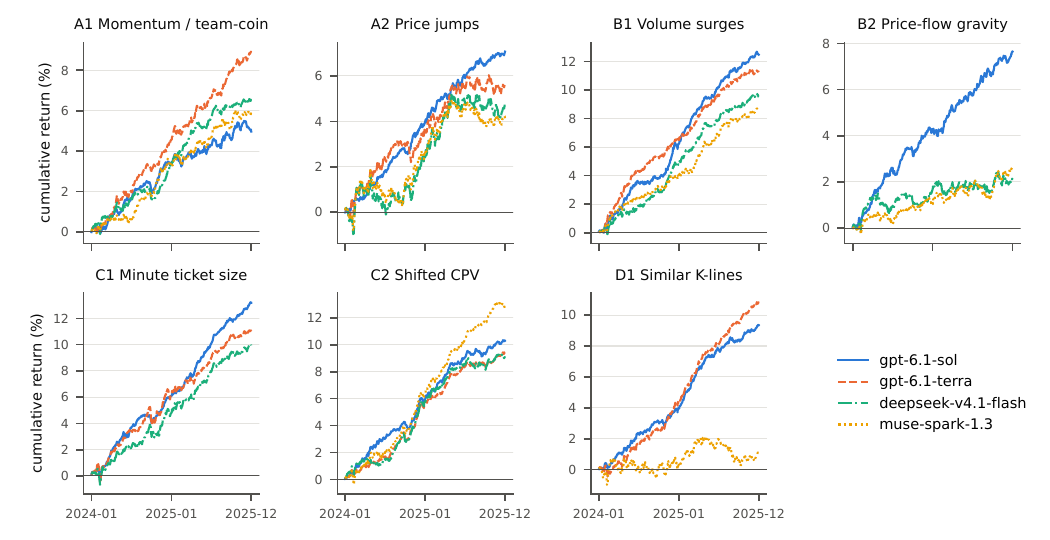}
  \caption{Cumulative returns (percent) of the factor pools on every task in the test period (2024-01-02 to 2025-12-31), i.e.\ the running sum of the style-free daily returns used in scoring, not the net value of a portfolio. The daily return is the style-free return of the top five groups over the average of all ten (the $x_d$ used in scoring), summed without transaction costs. Empty pools (B2-terra, C1-muse, D1-deepseek) have no curve; smoother and steeper curves mean higher Sharpe ratios.}
  \label{fig:cumulative}
\end{figure}
\subsection{Research Process Statistics}
\label{app:longhorizon}
Table~\ref{tab:longhorizon} gives statistics of the research process for each task.

\begin{table}[H]
  \centering
  \caption{Research process statistics per task: median (minimum--maximum over the 7 tasks in parentheses). Tool calls include command execution, terminal input and image viewing; tokens are the sum of per-request usage in the Codex session records.}
  \label{tab:longhorizon}
  \small
  \setlength{\tabcolsep}{4pt}
  \resizebox{\linewidth}{!}{\begin{tabular}{@{}lcccc@{}}
    \toprule
    & gpt-6.1-sol & gpt-6.1-terra & deepseek-v4.1-flash & muse-spark-1.3 \\
    \midrule
    Research time (hours) & 6.9 (5.8--7.8) & 2.0 (1.6--2.7) & 4.8 (3.3--6.0) & 4.8 (2.9--7.9) \\
    Tool calls & 387 (343--429) & 296 (269--402) & 1210 (832--1431) & 862 (763--1773) \\
    Context compactions & 4 (4--4) & 0 (0--1) & 0 (0--1) & 4 (3--14) \\
    Total hypotheses proposed & 152 (142--217) & 97 (71--135) & 103 (60--175) & 210 (101--270) \\
    Mean hypotheses in round 0 / round 6 & 40.0 / 15.4 & 22.4 / 11.3 & 40.7 / 6.0 & 49.1 / 11.4 \\
    Factors in pool & 6 (2--11) & 4 (0--7) & 3 (0--6) & 1 (0--10) \\
    Factors in own selection & 2 (1--3) & 2 (1--3) & 1 (1--2) & 1 (0--2) \\
    Output tokens (millions) & 0.65 (0.62--0.71) & 0.22 (0.18--0.26) & 1.36 (0.91--1.84) & 0.71 (0.57--1.49) \\
    Input tokens (millions) & 68 (56--78) & 30 (25--48) & 198 (81--294) & 101 (77--238) \\
    \bottomrule
  \end{tabular}}
\end{table}

\end{document}